\documentclass[twocolumn,showpacs,preprintnumbers,amsmath,amssymb,prl,unsortedaddress,superscriptaddress,longbibliography,letterpaper]{revtex4-1}
\usepackage{graphicx}% Include figure files
\usepackage[english]{babel}
\begin{document}

%\preprint{APS/123-QED}

\title{Quantum control via a genetic algorithm of the field ionization pathway of a Rydberg electron}

\author{Vincent C. Gregoric}%
\affiliation{Department of Physics, Bryn Mawr College, Bryn Mawr, PA 19010.}

\author{Xinyue Kang}
\affiliation{Department of Physics and Astronomy, Ursinus College, Collegeville, PA 19426.}

\author{Zhimin Cheryl Liu}%
\affiliation{Department of Physics, Bryn Mawr College, Bryn Mawr, PA 19010.}

\author{Zoe A. Rowley}
\affiliation{Department of Physics and Astronomy, Ursinus College, Collegeville, PA 19426.}

\author{Thomas J. Carroll}
\affiliation{Department of Physics and Astronomy, Ursinus College, Collegeville, PA 19426.}

\author{Michael W. Noel}%
\affiliation{Department of Physics, Bryn Mawr College, Bryn Mawr, PA 19010.}

\date{\today}% It is always \today, today,

\begin{abstract}
Quantum control of the pathway along which a Rydberg electron field ionizes is experimentally and computationally demonstrated. Selective field ionization is typically done with a slowly rising electric field pulse. The $(1/n^*)^4$ scaling of the classical ionization threshold leads to a rough mapping between arrival time of the electron signal and principal quantum number of the Rydberg electron. This is complicated by the many avoided level crossings that the electron must traverse on the way to ionization, which in general leads to broadening of the time-resolved field ionization signal. In order to control the ionization pathway, thus directing the signal to the desired arrival time, a perturbing electric field produced by an arbitrary waveform generator is added to a slowly rising electric field. A genetic algorithm evolves the perturbing field in an effort to achieve the target time-resolved field ionization signal. 
\end{abstract}

%\pacs{32.80.Ee, 32.60.+i}% PACS, the Physics and Astronomy

\maketitle

The study of quantum mechanics is motivated not only by the desire to understand microscopic phenomena, but also to control such systems. The field of quantum control offers a promising range of applications, from laser-controlled chemical reactions to quantum computing~\cite{brif_control_2010,saffman_quantum_2016,ladd_quantum_2010}. While there are many techniques available, they all rely on phase manipulation and coherence (i.e., interference effects) to control the system. 

One method that has been successfully used to implement quantum control is the genetic algorithm (GA), a stochastic optimization technique based on the tenets of Darwinian evolution~\cite{holland_adaptation_1992}. The use of GAs for quantum control typically involves tailoring the frequency, intensity, and phase of a laser pulse in order to achieve a desired effect. Judson and Rabitz first proposed a method to use a GA along with experimental feedback to control the laser excitation of molecules \textit{in situ}~\cite{judson_teaching_1992}. Since then, GAs have been used for a variety of theoretical and experimental applications in quantum control. Examples include improving the excitation efficiency in laser dye~\cite{bardeen_feedback_1997}, controlling the fragmentation of $\mathrm{S_8}$~\cite{wells_closed-loop_2005}, increasing the storage time of an EIT signal in a rare-earth-ion-doped crystal~\cite{heinze_stopped_2013}, manipulating the output spectrum in high-harmonic generation experiments~\cite{bartels_shaped-pulse_2000, chipperfield_ideal_2009}, selectively exciting vibrational states in molecular liquids~\cite{pearson_coherent_2001}, and designing NMR pulse sequences for state preparation and quantum gate operations~\cite{manu_singlet-state_2012}. 

The use of GAs is not limited to the field of quantum control; GAs have been applied to a diverse set of problems across multiple disciplines. Chemical physicists have used GAs to predict stable crystal and molecular structures by searching for low-energy configurations~\cite{oganov_crystal_2006, deaven_molecular_1995, alexandrova_structure_2004}. In microscopy, GAs have been used in combination with adaptive optic elements to reduce both off-axis~\cite{albert_smart_2000} and axial~\cite{wright_exploration_2005} aberrations. GAs have also been used to automate the fitting of spectroscopic data~\cite{hageman_direct_2000}, design optical resonators~\cite{gondarenko_spontaneous_2006}, minimize fluid drag~\cite{milano_clustering_2002}, and model insect flight~\cite{berman_energy-minimizing_2007}.

One system of interest for quantum control is a collection of atoms excited to Rydberg states of high principal quantum number $n$. Shaped laser pulses have been used to excite coherent superpositions of Rydberg states, thus controlling the shape of the electron wavefunction~\cite{weinacht_measurement_1998,pisharody_phase-controlled_2002,verlet_controlling_2002,noel_shaping_1997}. Alternatively, wavepackets can be excited or coherently manipulated with short electric field pulses~\cite{arbo_pulse-induced_2003,mestayer_transporting_2007} or microwave fields~\cite{maeda_microwave_2005}. Recently, quantum beats between different $\left|m_j\right|$ sublevels of the same fine structure state have been observed in the field ionization signal of Rydberg atoms~\cite{feynman_quantum_2015}. The presence of these quantum beats requires coherence throughout the ionization process. This motivated us to explore possible avenues for quantum control during field ionization. While the Hamiltonian describing the essential physics, the Stark effect, is well known for this system, the complexity of the ionization process makes this system of particular interest for exploring quantum control.

Since the valence electron in a Rydberg atom is weakly bound to the ion core, it is easily ionized by an external electric field, with higher energy states ionizing at lower fields. Selective field ionization (SFI) is a detection technique that takes advantage of this fact to gain insight into the state distribution of a group of Rydberg atoms~\cite{gallagher_rydberg_1994}. By increasing an external electric field gradually, the time at which the ionized electron is detected can be correlated to its initial state.

\begin{figure*}
	\centering
	\includegraphics{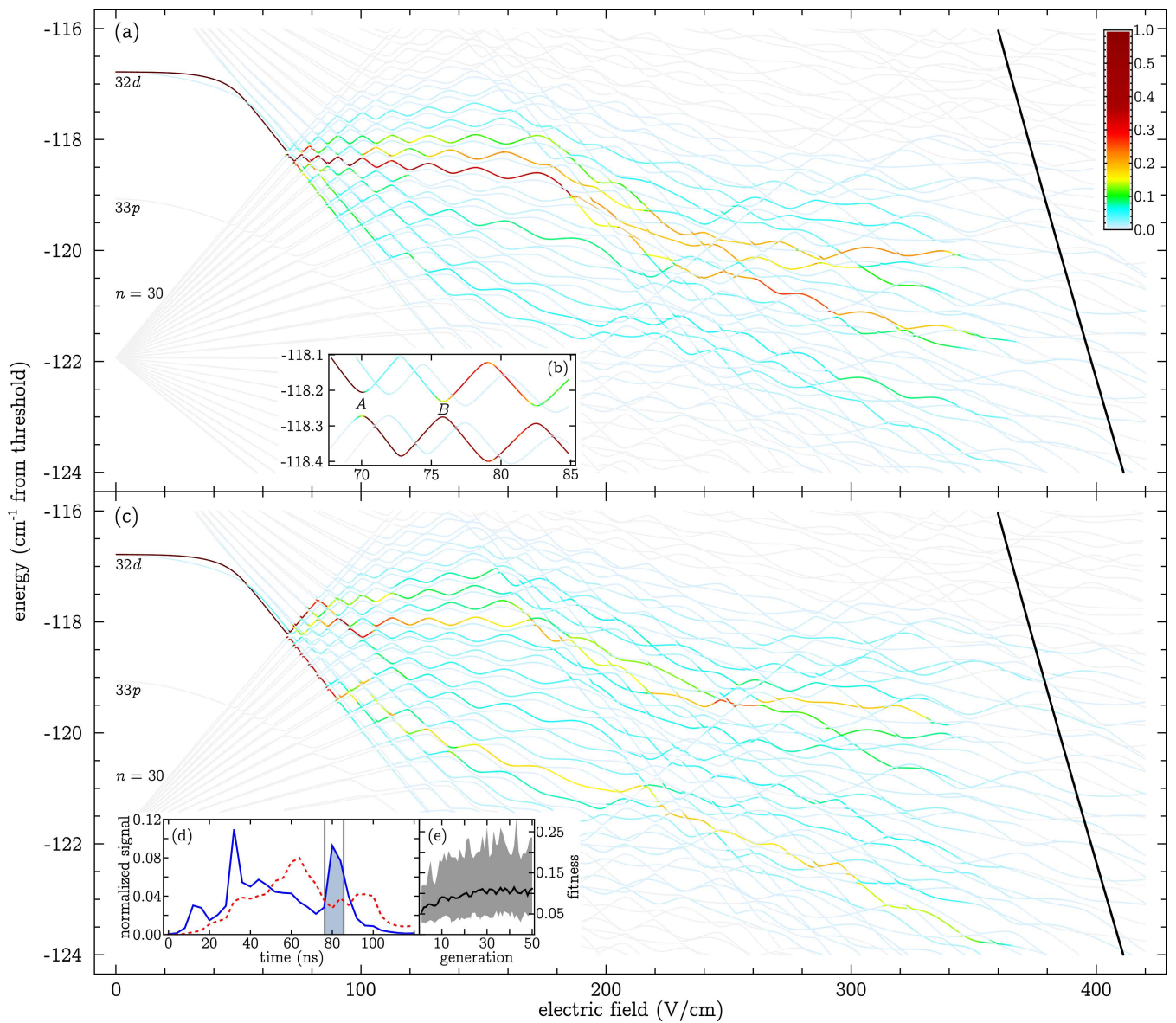}
	\caption{(Color online) Calculated Stark map showing the path to ionization for an electron initially excited to the $32d_{5/2,\left|m_j\right|=3/2}$ state of Rb$^{85}$. The classical ionization threshold is shown by the thick black line. The color of each line is determined by the electron amplitude. The calculation is fully quantum and proceeds as in Feynman \emph{et al.}~\cite{feynman_quantum_2015} with a basis including all states from the $n=29$ to the $n=33$ manifold, and a time resolution of 0.01~ns. (a) The path for the unperturbed ionizing electric field pulse. (b) Detail at the first few avoided crossings after the $32d$ lines hit the $n=30$ manifold. (c) The path for the best scoring optimized pulse from a simulation of the genetic algorithm. The calculation was performed in parallel for 48 pulses over 50 generations using the same GA as the experiment. (d) The calculated time resolved signal for both the optimized (solid blue) and unperturbed (dashed red) pulses. The gate region is highlighted. (e) Fitness score for the simulation. 
	}
	\label{fig:starkmap}
\end{figure*}

SFI is somewhat complicated by the Stark effect, which shifts the energy levels of Rydberg states in an external electric field. This leads to many avoided level crossings that the electron must traverse on the way to ionization (see Fig.~\ref{fig:starkmap}). Early studies in sodium near $n=32$ identified well-defined features associated with two ionization pathways, one predominantly adiabatic and one predominantly diabatic~\cite{jeys_diabatic_1980}. The ionization pathway is in general more complicated. At $n=32$, a Rydberg state will encounter several hundred avoided crossings on its way to ionization. An electron that begins in a single state at low field will spread out over multiple states as it passes through the avoided crossings, resulting in a broadened SFI signal. States that are closely separated in energy at low field are often unresolvable in the final signal~\cite{forre_selective-field-ionization_2003,feynman_quantum_2015}. The calculated path to ionization is shown in Fig.~\ref{fig:starkmap}(a) for the case of a roughly linear SFI pulse and in Fig.~\ref{fig:starkmap}(c) for an optimized pulse.

According to the Landau-Zener approximation, population transfer through an avoided crossing depends on the slew rate as well as the characteristics of the crossing~\cite{zener_non-adiabatic_1932,landau_zur_1932,stueckelberg_theorie_1932,majorana_atomi_1932}. Several groups have manipulated the slew rate with the goal of controlling the SFI signal shape. Tada \emph{et al.} used a field pulse with two sections of constant slew rate (first slow, then fast) to separate the signals from the $111s$ and $111p$ states in rubidium~\cite{tada_manipulating_2002}. In a similar experiment, G\"urtler and van der Zande used a pulse that increased quickly at first and then slowly in an effort to separate the $32s$, $31p$, and $30d$ states in rubidium~\cite{gurtler_-state_2004}. While they were not able to completely separate the signal from these three states, there was sufficient separation to determine the state distribution from an unknown superposition using the individual time-resolved signals as a basis set.

In this article, we present a demonstration of directed field ionization (DFI), a modification of SFI that allows one to coherently control the time-resolved field ionization signal through manipulation of the ionization pathway. This is achieved by using a GA to tailor the shape of the ionizing electric field pulse. We are able to not only change the magnitude of the slew rate, but also the sign, allowing us to traverse regions of the Stark map multiple times. For repeated traversals of the same avoided crossing or the complicated splitting and recombining shown in Fig.~\ref{fig:starkmap}(a) and (c), the interference of the relative phases of different pathways allows us to exert considerable control over the time resolved ionization signal.

Consider Fig.~\ref{fig:starkmap}(b), which shows the series of avoided crossings immediately after the states that are adiabatically connected to 32$d$ hit the $n=30$ manifold. When the electron traverses the first avoided crossing (labeled \emph{A}), a coherent superposition is created. As predicted by the Landau-Zener approximation, the traversal of the first avoided crossing is mostly diabatic, with only a small amplitude remaining in the upper state. The subsequent avoided crossing labeled \emph{B} is of similar size and would also be traversed diabatically except for the relative phase accumulated between \emph{A} and \emph{B}, which causes the amplitude to split more evenly. Because DFI relies on this interference, it is useful to think of the avoided crossings as beamsplitters for the electron wavefunction~\cite{feynman_quantum_2015}. Just as the output of an optical interferometer depends on the interference between two electromagnetic waves at a beamsplitter, the path of the electron is determined by the interference of the electron's wavefunction at an avoided crossing.  For a typical field ionization slew rate of 0.5 (V/cm)/ns, the relative phase accumulation between states in moving from $A$ to $B$ is $\approx$$100 \pi$.  In this case a change in slew rate of only 1\% will produce a $\pi$ shift in the accumulated phase.

Given the complexity of the path to ionization, there is no efficient analytical method for designing a desired pulse. A GA, however, requires only a target field ionization signal to evolve a solution. Our experimental setup allows for an upper limit of  $10^{1260}$ possible electric field pulses. While not all of these pulses will be measurably different, the solution space is clearly far too large for an exhaustive search. GAs also excel at searching large spaces for good, if not provably optimal, solutions. Finally, GAs are robust with respect to experimental conditions, which are automatically included in the optimization. For a more thorough overview of GAs see Mitchell~\cite{mitchell_introduction_1998} or Goldberg~\cite{goldberg_genetic_1989}.

\begin{figure}
	\includegraphics{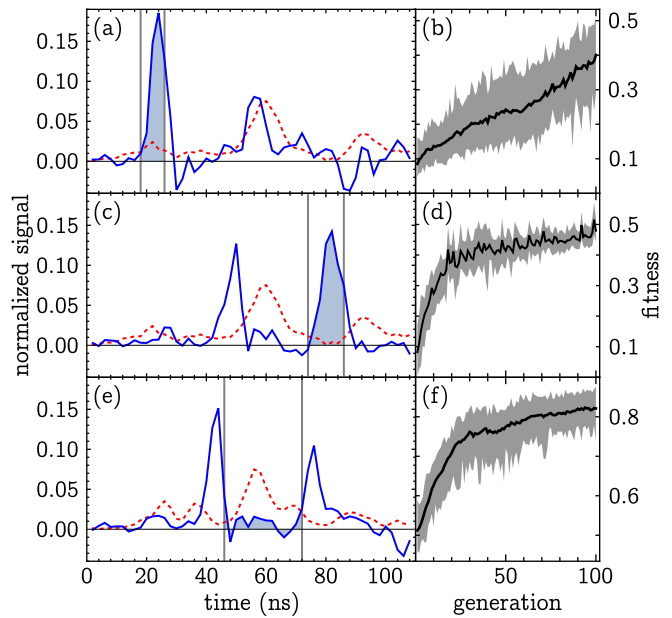}
	\caption{(Color online) Experimental results from three different DFI experiments. The unperturbed normalized ionization signal (dashed, red), the signal with the best fitness after 100 generations of evolution by GA (solid, blue), and the target gate (shaded blue region) are shown in (a), (c), and (e). The corresponding average fitness score (black) and range of fitness scores (shaded gray region) are shown in (b), (d), and (f), respectively. For (a) -- (d), the fitness score measures the percent of the signal within the target gate, while for (e) and (f), the fitness score measures the percent of the signal outside the target gate. 
	}
	\label{fig:evolvedtraces}
\end{figure}

A GA iteratively improves a population of potential solutions by evaluating them against a target and giving the characteristics of more successful candidates greater representation in subsequent generations. Our GA starts with a randomly generated population of 100 pulses, each consisting of 300 voltage values or \textit{genes}. We collect 10 shots for each pulse, interleaving the data collection, in order to average the field ionization signal. A \textit{fitness score} is calculated from the averaged signal based on how well it achieves the target solution; examples of different fitness scores can be seen in Fig.~\ref{fig:evolvedtraces}. The pulses are then ranked by fitness score and some number of the highest scoring pulses are propagated directly into the next generation, a technique known as \textit{elitism}. 

The majority of the next generation is created using \textit{crossover}, in which two parent pulses are mated together to produce a new child pulse. We select our parent pulses using \textit{tournament selection}. Two subsets of the population are randomly selected and the best scoring unique member of each subset is chosen as a parent. Then, for each gene locus of the child pulse, we randomly select one of the parent genes at the same locus. We repeat this process to fill the next generation's population.

Finally, we mutate the population by randomly selecting genes and assigning those genes a new random value. Mutation improves genetic diversity, which leads to a wider search of the solution space and prevents premature convergence to a local optimum. However, mutation can also destroy good solutions. We therefore use a low probability of mutation that is dynamically decreased. The entire algorithm is repeated until a fixed number of generations has passed.

Our experiment is done in a fairly standard magneto-optical trap containing $\approx$$10^6$ Rb$^{85}$ atoms at $\approx$200~$\mu$K. The excitation to Rydberg states is achieved using homemade external-cavity diode lasers~\cite{fahey_excitation_2011}. The 780~nm trapping laser excites the Rb atoms from the $5s_{1/2}$ ground state to the $5p_{3/2}$ state. A 10~$\mu$s wide 776~nm laser pulse drives the $5p_{3/2}\rightarrow5d_{5/2}$ transition, and the atoms are allowed to radiatively decay to the $6p_{3/2}$ state. The final excitation step to the $32d_{5/2, \left|m_j\right|=3/2}$ state is provided by a 1~$\mu$s wide 1022~nm laser pulse. All lasers are frequency stabilized using homemade electronic feedback circuits and either saturated absorption spectroscopy (780~nm and 776~nm) or an actively stabilized Fabry-P\'erot cavity (1022~nm). The continuous-wave output beams from the 776~nm and 1022~nm lasers are pulsed using acousto-optic modulators.

After excitation to Rydberg states, we apply a roughly linear ionizing electric field ramp using a trigger-transformer circuit, as well as a perturbing electric field (determined by the GA) from an arbitrary waveform generator. The trapped Rydberg atoms are located on-axis midway between two cylindrical electrodes which control the electric fields inside our vacuum chamber. A detailed discussion of the electrode geometry can be found in Fahey \emph{et al.}~\cite{fahey_imaging_2015}. The linear field ramp is applied to one of the cylinders, rising to $\approx$600~V/cm in $\approx$1.5~$\mu$s. Simultaneously, the perturbing electric field from the arbitrary waveform generator is applied to the other cylinder. The electrons arrive at the detector roughly 10~ns after ionization, where the time-resolved signal is amplified with a pair of multichannel plates.

Our waveform generator has 14-bit resolution, a sample rate of 1~GS/s, and can switch between extreme values of $\pm$10~V (corresponding to fields of $\pm3.8$~V/cm) in 3.3~ns. Given our electrode geometry and the available voltages, we are able to access electric field slew rates ranging from $-$1.6 to 3.0~(V/cm)/ns during ionization, allowing us to sweep through a typical avoided crossing three times. In the context of Fig.~\ref{fig:starkmap}(b), our waveform generator is capable of a minimum phase adjustment of $\approx$$\pi/20$ as the field rises from $A$ to $B$.

In addition to the fields from the trigger-transformer and the arbitrary waveform generator, we also apply a DC electric field of $\approx$6~V/cm to one of the cylinders to break the degeneracy between the $\left|m_j\right|~=~1/2$, 3/2, and 5/2 states. This allows us to excite a single $\left|m_j\right|$ state. The total electric field experienced by the atoms is the vector sum of these three fields.

Using this experimental setup, we began by verifying the coherence of the ionization process by repeating the interference experiment described in Feynman \emph{et al.}~\cite{feynman_quantum_2015}. After coherence was confirmed, we tested the ability of our GA to control the shape of the ionization signal \textit{in situ}. Appropriate values for the GA parameters (e.g. mutation rate, tournament size, etc.) were chosen both by simulating our GA and by trial and error.

The results for three different DFI experiments are shown in Fig.~\ref{fig:evolvedtraces}. In each case, we compare the result due to the unperturbed linear ramp with the result due to the optimized perturbed field. Since the signal level fluctuated on both short and long time scales, we normalized the area under the ionization signal for each shot of the experiment. In Fig.~\ref{fig:evolvedtraces}(a) and (b), the goal was to shift the signal into the leading edge of the unperturbed SFI signal. The fitness score was calculated by dividing the signal within the desired gate by the total signal. Initially, $8.3\%$ of the signal was within the target gate. After optimization, this was increased to $50.6\%$ for the best performing member of the final generation. For Fig.~\ref{fig:evolvedtraces}(c) and (d), the target gate was shifted later in time to a region which initially had very little signal. The GA optimization procedure was able to increase the signal within this gate from $4.4\%$ to $52.0\%$. The goal for the DFI experiment shown in Fig.~\ref{fig:evolvedtraces}(e) and (f) was to push signal out of the target gate. The amount of signal \textit{outside} the target gate increased from $50.4\%$ to $87.7\%$ as a result of optimization. We ran each optimization many times over the course of several weeks. The evolution proved to be very robust even in the presence of large signal fluctuations.

% (a): 8.5 +/- 1.9%    ---->    40.1 +/- 4.8%   (50.6% max)        (8.3% in the red trace scan)
% (b): 8.2 +/- 3.5%    ---->    47.9 +/- 1.6%   (52.0% max)        (4.4% in the red trace scan)
% (c): 51.2 +/- 2.6%   ---->    82.1 +/- 2.0%   (87.7% max)        (50.4% in the red trace scan)

One question that naturally arises is whether we can glean any physical insight by looking at the optimized ionization pulses. With this goal in mind, we have taken 20 datasets with identical parameters. When comparing the best performing arbitrary waveforms, we find that the region with the most similarities occurs during a 200~ns period just before and during ionization. Some of the signal shifts may be due to the more complicated temporal dependence of the electric field during ionization. Given the limitations of our trigger-transformer and arbitrary waveform generator, such shifts are limited to 8~ns, which are significantly smaller than the shifts seen in Fig.~\ref{fig:evolvedtraces}. It is clear that this effect cannot account for the entire optimization. To further confirm the role of coherent control in our optimization, we have taken a series of datasets with a range of arbitrary waveform end times (leaving the start time and all other parameters identical). Although the fitness score improves more dramatically when the arbitrary waveform is allowed to extend through ionization, there is still significant improvement even when the arbitrary waveform ends well before ionization. In fact, significant optimization is possible when the perturbing waveform is limited to a 100~ns region during the $1.5~\mu$s ionization pulse. While these experiments have given us insight into the important features of optimization, we cannot see the full ionization pathway experimentally.

In order to examine the ionization pathway, we have simulated the evolution of a population of 48 pulses over 50 generations for the gate shown in Fig.~\ref{fig:evolvedtraces}(c). The calculated path to ionization for the best scoring optimized pulse is shown in Fig.~\ref{fig:starkmap}(c), along with the time resolved signal and fitness score in (d) and (e). In early generations the GA can easily improve the fitness score by moving population from states that ionize just outside of the gate to neighboring states that ionize within the gate. These optimizations may inadvertently redistribute population among more distant states, which has no direct effect on the fitness score. This could have a negative effect on subsequent evolution by moving population that ionizes far from the gate even farther away.

We observe this effect for the optimized field ionization signals in both experiment and simulation, shown in Fig.~\ref{fig:evolvedtraces}(c) and Fig.~\ref{fig:starkmap}(d) respectively. Both display a similar double-peaked structure that results from the path to ionization shown in Fig.~\ref{fig:starkmap}(c), where the optimizations also send some population to earlier ionization. We are exploring alternative fitness scores to improve performance.

In conclusion, we have demonstrated the use of a GA to control the shape of the time-resolved field ionization signal \textit{in situ}. While we have seen a dramatic improvement in fitness score in each case, we have not been able to move \emph{all} of the signal into the desired gate. Improvements might be possible with different values for the GA parameters. Additionally, switching to a higher-$n$ state would allow for a slower slew rate for the unperturbed SFI pulse, increasing our ability to perturb the slew rate with the arbitrary waveform generator. We are in the process of investigating both of these options for improvement. We are also exploring the possibility of using DFI to separate previously unresolvable states in the ionization signal. 

This work was supported by the National Science Foundation under Grants No. 1607335 and No. 1607377 and the Howard Hughes Medical Institute. We also used the Extreme Science and Engineering Discovery Environment (XSEDE), which is supported by National Science Foundation grant number OCI-1053575.%\bibliographystyle{apsrev4-1}
%\bibliography{directedFieldIonization}% Produces the bibliography via BibTeX.
\bibliography{biblioFromZotero}% Produces the bibliography via BibTeX.

\begin{thebibliography}{42}%
\makeatletter
\providecommand \@ifxundefined [1]{%
 \@ifx{#1\undefined}
}%
\providecommand \@ifnum [1]{%
 \ifnum #1\expandafter \@firstoftwo
 \else \expandafter \@secondoftwo
 \fi
}%
\providecommand \@ifx [1]{%
 \ifx #1\expandafter \@firstoftwo
 \else \expandafter \@secondoftwo
 \fi
}%
\providecommand \natexlab [1]{#1}%
\providecommand \enquote  [1]{``#1''}%
\providecommand \bibnamefont  [1]{#1}%
\providecommand \bibfnamefont [1]{#1}%
\providecommand \citenamefont [1]{#1}%
\providecommand \href@noop [0]{\@secondoftwo}%
\providecommand \href [0]{\begingroup \@sanitize@url \@href}%
\providecommand \@href[1]{\@@startlink{#1}\@@href}%
\providecommand \@@href[1]{\endgroup#1\@@endlink}%
\providecommand \@sanitize@url [0]{\catcode `\\12\catcode `\$12\catcode
  `\&12\catcode `\#12\catcode `\^12\catcode `\_12\catcode `\%12\relax}%
\providecommand \@@startlink[1]{}%
\providecommand \@@endlink[0]{}%
\providecommand \url  [0]{\begingroup\@sanitize@url \@url }%
\providecommand \@url [1]{\endgroup\@href {#1}{\urlprefix }}%
\providecommand \urlprefix  [0]{URL }%
\providecommand \Eprint [0]{\href }%
\providecommand \doibase [0]{http://dx.doi.org/}%
\providecommand \selectlanguage [0]{\@gobble}%
\providecommand \bibinfo  [0]{\@secondoftwo}%
\providecommand \bibfield  [0]{\@secondoftwo}%
\providecommand \translation [1]{[#1]}%
\providecommand \BibitemOpen [0]{}%
\providecommand \bibitemStop [0]{}%
\providecommand \bibitemNoStop [0]{.\EOS\space}%
\providecommand \EOS [0]{\spacefactor3000\relax}%
\providecommand \BibitemShut  [1]{\csname bibitem#1\endcsname}%
\let\auto@bib@innerbib\@empty
%</preamble>
\bibitem [{\citenamefont {Brif}\ \emph {et~al.}(2010)\citenamefont {Brif},
  \citenamefont {Chakrabarti},\ and\ \citenamefont
  {Rabitz}}]{brif_control_2010}%
  \BibitemOpen
  \bibfield  {author} {\bibinfo {author} {\bibfnamefont {Constantin}\
  \bibnamefont {Brif}}, \bibinfo {author} {\bibfnamefont {Raj}\ \bibnamefont
  {Chakrabarti}}, \ and\ \bibinfo {author} {\bibfnamefont {Herschel}\
  \bibnamefont {Rabitz}},\ }\bibfield  {title} {\enquote {\bibinfo {title}
  {Control of quantum phenomena: past, present and future},}\ }\href {\doibase
  10.1088/1367-2630/12/7/075008} {\bibfield  {journal} {\bibinfo  {journal}
  {New J. Phys.}\ }\textbf {\bibinfo {volume} {12}},\ \bibinfo {pages} {075008}
  (\bibinfo {year} {2010})}\BibitemShut {NoStop}%
\bibitem [{\citenamefont {Saffman}(2016)}]{saffman_quantum_2016}%
  \BibitemOpen
  \bibfield  {author} {\bibinfo {author} {\bibfnamefont {M.}~\bibnamefont
  {Saffman}},\ }\bibfield  {title} {\enquote {\bibinfo {title} {Quantum
  computing with atomic qubits and {Rydberg} interactions: progress and
  challenges},}\ }\href {\doibase 10.1088/0953-4075/49/20/202001} {\bibfield
  {journal} {\bibinfo  {journal} {J. Phys. B: At. Mol. Opt. Phys.}\ }\textbf
  {\bibinfo {volume} {49}},\ \bibinfo {pages} {202001} (\bibinfo {year}
  {2016})}\BibitemShut {NoStop}%
\bibitem [{\citenamefont {Ladd}\ \emph {et~al.}(2010)\citenamefont {Ladd},
  \citenamefont {Jelezko}, \citenamefont {Laflamme}, \citenamefont {Nakamura},
  \citenamefont {Monroe},\ and\ \citenamefont
  {O{\textquoteright}Brien}}]{ladd_quantum_2010}%
  \BibitemOpen
  \bibfield  {author} {\bibinfo {author} {\bibfnamefont {T.~D.}\ \bibnamefont
  {Ladd}}, \bibinfo {author} {\bibfnamefont {F.}~\bibnamefont {Jelezko}},
  \bibinfo {author} {\bibfnamefont {R.}~\bibnamefont {Laflamme}}, \bibinfo
  {author} {\bibfnamefont {Y.}~\bibnamefont {Nakamura}}, \bibinfo {author}
  {\bibfnamefont {C.}~\bibnamefont {Monroe}}, \ and\ \bibinfo {author}
  {\bibfnamefont {J.~L.}\ \bibnamefont {O{\textquoteright}Brien}},\ }\bibfield
  {title} {\enquote {\bibinfo {title} {Quantum computers},}\ }\href {\doibase
  10.1038/nature08812} {\bibfield  {journal} {\bibinfo  {journal} {Nature}\
  }\textbf {\bibinfo {volume} {464}},\ \bibinfo {pages} {45} (\bibinfo {year}
  {2010})}\BibitemShut {NoStop}%
\bibitem [{\citenamefont {Holland}(1992)}]{holland_adaptation_1992}%
  \BibitemOpen
  \bibfield  {author} {\bibinfo {author} {\bibfnamefont {John~H.}\ \bibnamefont
  {Holland}},\ }\href@noop {} {\emph {\bibinfo {title} {Adaptation in {Natural}
  and {Artificial} {Systems}: {An} {Introductory} {Analysis} with
  {Applications} to {Biology}, {Control}, and {Artificial} {Intelligence}}}},\
  \bibinfo {edition} {reprint edition}\ ed.\ (\bibinfo  {publisher} {A Bradford
  Book},\ \bibinfo {address} {Cambridge, Mass},\ \bibinfo {year}
  {1992})\BibitemShut {NoStop}%
\bibitem [{\citenamefont {Judson}\ and\ \citenamefont
  {Rabitz}(1992)}]{judson_teaching_1992}%
  \BibitemOpen
  \bibfield  {author} {\bibinfo {author} {\bibfnamefont {Richard~S.}\
  \bibnamefont {Judson}}\ and\ \bibinfo {author} {\bibfnamefont {Herschel}\
  \bibnamefont {Rabitz}},\ }\bibfield  {title} {\enquote {\bibinfo {title}
  {Teaching lasers to control molecules},}\ }\href {\doibase
  10.1103/PhysRevLett.68.1500} {\bibfield  {journal} {\bibinfo  {journal}
  {Phys. Rev. Lett.}\ }\textbf {\bibinfo {volume} {68}},\ \bibinfo {pages}
  {1500} (\bibinfo {year} {1992})}\BibitemShut {NoStop}%
\bibitem [{\citenamefont {Bardeen}\ \emph {et~al.}(1997)\citenamefont
  {Bardeen}, \citenamefont {Yakovlev}, \citenamefont {Wilson}, \citenamefont
  {Carpenter}, \citenamefont {Weber},\ and\ \citenamefont
  {Warren}}]{bardeen_feedback_1997}%
  \BibitemOpen
  \bibfield  {author} {\bibinfo {author} {\bibfnamefont {Christopher~J}\
  \bibnamefont {Bardeen}}, \bibinfo {author} {\bibfnamefont {Vladislav~V}\
  \bibnamefont {Yakovlev}}, \bibinfo {author} {\bibfnamefont {Kent~R}\
  \bibnamefont {Wilson}}, \bibinfo {author} {\bibfnamefont {Scott~D}\
  \bibnamefont {Carpenter}}, \bibinfo {author} {\bibfnamefont {Peter~M}\
  \bibnamefont {Weber}}, \ and\ \bibinfo {author} {\bibfnamefont {Warren~S}\
  \bibnamefont {Warren}},\ }\bibfield  {title} {\enquote {\bibinfo {title}
  {Feedback quantum control of molecular electronic population transfer},}\
  }\href {\doibase 10.1016/S0009-2614(97)01081-6} {\bibfield  {journal}
  {\bibinfo  {journal} {Chem. Phys. Lett.}\ }\textbf {\bibinfo {volume}
  {280}},\ \bibinfo {pages} {151} (\bibinfo {year} {1997})}\BibitemShut
  {NoStop}%
\bibitem [{\citenamefont {Wells}\ \emph {et~al.}(2005)\citenamefont {Wells},
  \citenamefont {Betsch}, \citenamefont {Conover}, \citenamefont {DeWitt},
  \citenamefont {Pinkham},\ and\ \citenamefont
  {Jones}}]{wells_closed-loop_2005}%
  \BibitemOpen
  \bibfield  {author} {\bibinfo {author} {\bibfnamefont {E.}~\bibnamefont
  {Wells}}, \bibinfo {author} {\bibfnamefont {K.~J.}\ \bibnamefont {Betsch}},
  \bibinfo {author} {\bibfnamefont {C.~W.~S.}\ \bibnamefont {Conover}},
  \bibinfo {author} {\bibfnamefont {Merrick~J.}\ \bibnamefont {DeWitt}},
  \bibinfo {author} {\bibfnamefont {D.}~\bibnamefont {Pinkham}}, \ and\
  \bibinfo {author} {\bibfnamefont {R.~R.}\ \bibnamefont {Jones}},\ }\bibfield
  {title} {\enquote {\bibinfo {title} {Closed-loop control of intense-laser
  fragmentation of $\mathrm{S}_{8}$},}\ }\href {\doibase
  10.1103/PhysRevA.72.063406} {\bibfield  {journal} {\bibinfo  {journal} {Phys.
  Rev. A}\ }\textbf {\bibinfo {volume} {72}},\ \bibinfo {pages} {063406}
  (\bibinfo {year} {2005})}\BibitemShut {NoStop}%
\bibitem [{\citenamefont {Heinze}\ \emph {et~al.}(2013)\citenamefont {Heinze},
  \citenamefont {Hubrich},\ and\ \citenamefont
  {Halfmann}}]{heinze_stopped_2013}%
  \BibitemOpen
  \bibfield  {author} {\bibinfo {author} {\bibfnamefont {Georg}\ \bibnamefont
  {Heinze}}, \bibinfo {author} {\bibfnamefont {Christian}\ \bibnamefont
  {Hubrich}}, \ and\ \bibinfo {author} {\bibfnamefont {Thomas}\ \bibnamefont
  {Halfmann}},\ }\bibfield  {title} {\enquote {\bibinfo {title} {Stopped
  {Light} and {Image} {Storage} by {Electromagnetically} {Induced}
  {Transparency} up to the {Regime} of {One} {Minute}},}\ }\href {\doibase
  10.1103/PhysRevLett.111.033601} {\bibfield  {journal} {\bibinfo  {journal}
  {Phys. Rev. Lett.}\ }\textbf {\bibinfo {volume} {111}},\ \bibinfo {pages}
  {033601} (\bibinfo {year} {2013})}\BibitemShut {NoStop}%
\bibitem [{\citenamefont {Bartels}\ \emph {et~al.}(2000)\citenamefont
  {Bartels}, \citenamefont {Backus}, \citenamefont {Zeek}, \citenamefont
  {Misoguti}, \citenamefont {Vdovin}, \citenamefont {Christov}, \citenamefont
  {Murnane},\ and\ \citenamefont {Kapteyn}}]{bartels_shaped-pulse_2000}%
  \BibitemOpen
  \bibfield  {author} {\bibinfo {author} {\bibfnamefont {R.}~\bibnamefont
  {Bartels}}, \bibinfo {author} {\bibfnamefont {S.}~\bibnamefont {Backus}},
  \bibinfo {author} {\bibfnamefont {E.}~\bibnamefont {Zeek}}, \bibinfo {author}
  {\bibfnamefont {L.}~\bibnamefont {Misoguti}}, \bibinfo {author}
  {\bibfnamefont {G.}~\bibnamefont {Vdovin}}, \bibinfo {author} {\bibfnamefont
  {I.~P.}\ \bibnamefont {Christov}}, \bibinfo {author} {\bibfnamefont {M.~M.}\
  \bibnamefont {Murnane}}, \ and\ \bibinfo {author} {\bibfnamefont {H.~C.}\
  \bibnamefont {Kapteyn}},\ }\bibfield  {title} {{\selectlanguage
  {english}\enquote {\bibinfo {title} {Shaped-pulse optimization of coherent
  emission of high-harmonic soft {X}-rays},}\ }}\href {\doibase
  10.1038/35018029} {\bibfield  {journal} {\bibinfo  {journal} {Nature}\
  }\textbf {\bibinfo {volume} {406}},\ \bibinfo {pages} {164} (\bibinfo {year}
  {2000})}\BibitemShut {NoStop}%
\bibitem [{\citenamefont {Chipperfield}\ \emph {et~al.}(2009)\citenamefont
  {Chipperfield}, \citenamefont {Robinson}, \citenamefont {Tisch},\ and\
  \citenamefont {Marangos}}]{chipperfield_ideal_2009}%
  \BibitemOpen
  \bibfield  {author} {\bibinfo {author} {\bibfnamefont {L.~E.}\ \bibnamefont
  {Chipperfield}}, \bibinfo {author} {\bibfnamefont {J.~S.}\ \bibnamefont
  {Robinson}}, \bibinfo {author} {\bibfnamefont {J.~W.~G.}\ \bibnamefont
  {Tisch}}, \ and\ \bibinfo {author} {\bibfnamefont {J.~P.}\ \bibnamefont
  {Marangos}},\ }\bibfield  {title} {\enquote {\bibinfo {title} {Ideal
  {Waveform} to {Generate} the {Maximum} {Possible} {Electron} {Recollision}
  {Energy} for {Any} {Given} {Oscillation} {Period}},}\ }\href {\doibase
  10.1103/PhysRevLett.102.063003} {\bibfield  {journal} {\bibinfo  {journal}
  {Phys. Rev. Lett.}\ }\textbf {\bibinfo {volume} {102}},\ \bibinfo {pages}
  {063003} (\bibinfo {year} {2009})}\BibitemShut {NoStop}%
\bibitem [{\citenamefont {Pearson}\ \emph {et~al.}(2001)\citenamefont
  {Pearson}, \citenamefont {White}, \citenamefont {Weinacht},\ and\
  \citenamefont {Bucksbaum}}]{pearson_coherent_2001}%
  \BibitemOpen
  \bibfield  {author} {\bibinfo {author} {\bibfnamefont {B.~J.}\ \bibnamefont
  {Pearson}}, \bibinfo {author} {\bibfnamefont {J.~L.}\ \bibnamefont {White}},
  \bibinfo {author} {\bibfnamefont {T.~C.}\ \bibnamefont {Weinacht}}, \ and\
  \bibinfo {author} {\bibfnamefont {P.~H.}\ \bibnamefont {Bucksbaum}},\
  }\bibfield  {title} {\enquote {\bibinfo {title} {Coherent control using
  adaptive learning algorithms},}\ }\href {\doibase 10.1103/PhysRevA.63.063412}
  {\bibfield  {journal} {\bibinfo  {journal} {Phys. Rev. A}\ }\textbf {\bibinfo
  {volume} {63}},\ \bibinfo {pages} {063412} (\bibinfo {year}
  {2001})}\BibitemShut {NoStop}%
\bibitem [{\citenamefont {Manu}\ and\ \citenamefont
  {Kumar}(2012)}]{manu_singlet-state_2012}%
  \BibitemOpen
  \bibfield  {author} {\bibinfo {author} {\bibfnamefont {V.~S.}\ \bibnamefont
  {Manu}}\ and\ \bibinfo {author} {\bibfnamefont {Anil}\ \bibnamefont
  {Kumar}},\ }\bibfield  {title} {\enquote {\bibinfo {title} {Singlet-state
  creation and universal quantum computation in {NMR} using a genetic
  algorithm},}\ }\href {\doibase 10.1103/PhysRevA.86.022324} {\bibfield
  {journal} {\bibinfo  {journal} {Phys. Rev. A}\ }\textbf {\bibinfo {volume}
  {86}},\ \bibinfo {pages} {022324} (\bibinfo {year} {2012})}\BibitemShut
  {NoStop}%
\bibitem [{\citenamefont {Oganov}\ and\ \citenamefont
  {Glass}(2006)}]{oganov_crystal_2006}%
  \BibitemOpen
  \bibfield  {author} {\bibinfo {author} {\bibfnamefont {Artem~R.}\
  \bibnamefont {Oganov}}\ and\ \bibinfo {author} {\bibfnamefont {Colin~W.}\
  \bibnamefont {Glass}},\ }\bibfield  {title} {{\selectlanguage
  {english}\enquote {\bibinfo {title} {Crystal structure prediction using
  \textit{ab initio} evolutionary techniques: {Principles} and applications},}\
  }}\href {\doibase 10.1063/1.2210932} {\bibfield  {journal} {\bibinfo
  {journal} {J. Chem. Phys.}\ }\textbf {\bibinfo {volume} {124}},\ \bibinfo
  {pages} {244704} (\bibinfo {year} {2006})}\BibitemShut {NoStop}%
\bibitem [{\citenamefont {Deaven}\ and\ \citenamefont
  {Ho}(1995)}]{deaven_molecular_1995}%
  \BibitemOpen
  \bibfield  {author} {\bibinfo {author} {\bibfnamefont {D.~M.}\ \bibnamefont
  {Deaven}}\ and\ \bibinfo {author} {\bibfnamefont {K.~M.}\ \bibnamefont
  {Ho}},\ }\bibfield  {title} {\enquote {\bibinfo {title} {Molecular {Geometry}
  {Optimization} with a {Genetic} {Algorithm}},}\ }\href {\doibase
  10.1103/PhysRevLett.75.288} {\bibfield  {journal} {\bibinfo  {journal} {Phys.
  Rev. Lett.}\ }\textbf {\bibinfo {volume} {75}},\ \bibinfo {pages} {288}
  (\bibinfo {year} {1995})}\BibitemShut {NoStop}%
\bibitem [{\citenamefont {Alexandrova}\ \emph {et~al.}(2004)\citenamefont
  {Alexandrova}, \citenamefont {Boldyrev}, \citenamefont {Fu}, \citenamefont
  {Yang}, \citenamefont {Wang},\ and\ \citenamefont
  {Wang}}]{alexandrova_structure_2004}%
  \BibitemOpen
  \bibfield  {author} {\bibinfo {author} {\bibfnamefont {Anastassia~N.}\
  \bibnamefont {Alexandrova}}, \bibinfo {author} {\bibfnamefont {Alexander~I.}\
  \bibnamefont {Boldyrev}}, \bibinfo {author} {\bibfnamefont {You-Jun}\
  \bibnamefont {Fu}}, \bibinfo {author} {\bibfnamefont {Xin}\ \bibnamefont
  {Yang}}, \bibinfo {author} {\bibfnamefont {Xue-Bin}\ \bibnamefont {Wang}}, \
  and\ \bibinfo {author} {\bibfnamefont {Lai-Sheng}\ \bibnamefont {Wang}},\
  }\bibfield  {title} {{\selectlanguage {english}\enquote {\bibinfo {title}
  {Structure of the $\mathrm{Na}_x\mathrm{Cl}_{x+1}^- (x=1{\textendash}4)$
  clusters via \textit{ab initio} genetic algorithm and photoelectron
  spectroscopy},}\ }}\href {\doibase 10.1063/1.1783276} {\bibfield  {journal}
  {\bibinfo  {journal} {J. Chem. Phys.}\ }\textbf {\bibinfo {volume} {121}},\
  \bibinfo {pages} {5709} (\bibinfo {year} {2004})}\BibitemShut {NoStop}%
\bibitem [{\citenamefont {Albert}\ \emph {et~al.}(2000)\citenamefont {Albert},
  \citenamefont {Sherman}, \citenamefont {Mourou}, \citenamefont {Norris},\
  and\ \citenamefont {Vdovin}}]{albert_smart_2000}%
  \BibitemOpen
  \bibfield  {author} {\bibinfo {author} {\bibfnamefont {O.}~\bibnamefont
  {Albert}}, \bibinfo {author} {\bibfnamefont {L.}~\bibnamefont {Sherman}},
  \bibinfo {author} {\bibfnamefont {G.}~\bibnamefont {Mourou}}, \bibinfo
  {author} {\bibfnamefont {T.~B.}\ \bibnamefont {Norris}}, \ and\ \bibinfo
  {author} {\bibfnamefont {G.}~\bibnamefont {Vdovin}},\ }\bibfield  {title}
  {{\selectlanguage {english}\enquote {\bibinfo {title} {Smart microscope: an
  adaptive optics learning system for aberration correction in multiphoton
  confocal microscopy},}\ }}\href {\doibase 10.1364/OL.25.000052} {\bibfield
  {journal} {\bibinfo  {journal} {Opt. Lett.}\ }\textbf {\bibinfo {volume}
  {25}},\ \bibinfo {pages} {52} (\bibinfo {year} {2000})}\BibitemShut {NoStop}%
\bibitem [{\citenamefont {Wright}\ \emph {et~al.}(2005)\citenamefont {Wright},
  \citenamefont {Burns}, \citenamefont {Patterson}, \citenamefont {Poland},
  \citenamefont {Valentine},\ and\ \citenamefont
  {Girkin}}]{wright_exploration_2005}%
  \BibitemOpen
  \bibfield  {author} {\bibinfo {author} {\bibfnamefont {Amanda~J.}\
  \bibnamefont {Wright}}, \bibinfo {author} {\bibfnamefont {David}\
  \bibnamefont {Burns}}, \bibinfo {author} {\bibfnamefont {Brett~A.}\
  \bibnamefont {Patterson}}, \bibinfo {author} {\bibfnamefont {Simon~P.}\
  \bibnamefont {Poland}}, \bibinfo {author} {\bibfnamefont {Gareth~J.}\
  \bibnamefont {Valentine}}, \ and\ \bibinfo {author} {\bibfnamefont {John~M.}\
  \bibnamefont {Girkin}},\ }\bibfield  {title} {{\selectlanguage
  {english}\enquote {\bibinfo {title} {Exploration of the optimisation
  algorithms used in the implementation of adaptive optics in confocal and
  multiphoton microscopy},}\ }}\href {\doibase 10.1002/jemt.20178} {\bibfield
  {journal} {\bibinfo  {journal} {Microsc. Res. Tech.}\ }\textbf {\bibinfo
  {volume} {67}},\ \bibinfo {pages} {36} (\bibinfo {year} {2005})}\BibitemShut
  {NoStop}%
\bibitem [{\citenamefont {Hageman}\ \emph {et~al.}(2000)\citenamefont
  {Hageman}, \citenamefont {Wehrens}, \citenamefont {de~Gelder}, \citenamefont
  {Leo~Meerts},\ and\ \citenamefont {Buydens}}]{hageman_direct_2000}%
  \BibitemOpen
  \bibfield  {author} {\bibinfo {author} {\bibfnamefont {J.~A.}\ \bibnamefont
  {Hageman}}, \bibinfo {author} {\bibfnamefont {R.}~\bibnamefont {Wehrens}},
  \bibinfo {author} {\bibfnamefont {R.}~\bibnamefont {de~Gelder}}, \bibinfo
  {author} {\bibfnamefont {W.}~\bibnamefont {Leo~Meerts}}, \ and\ \bibinfo
  {author} {\bibfnamefont {L.~M.~C.}\ \bibnamefont {Buydens}},\ }\bibfield
  {title} {{\selectlanguage {english}\enquote {\bibinfo {title} {Direct
  determination of molecular constants from rovibronic spectra with genetic
  algorithms},}\ }}\href {\doibase 10.1063/1.1314353} {\bibfield  {journal}
  {\bibinfo  {journal} {J. Chem. Phys.}\ }\textbf {\bibinfo {volume} {113}},\
  \bibinfo {pages} {7955} (\bibinfo {year} {2000})}\BibitemShut {NoStop}%
\bibitem [{\citenamefont {Gondarenko}\ \emph {et~al.}(2006)\citenamefont
  {Gondarenko}, \citenamefont {Preble}, \citenamefont {Robinson}, \citenamefont
  {Chen}, \citenamefont {Lipson},\ and\ \citenamefont
  {Lipson}}]{gondarenko_spontaneous_2006}%
  \BibitemOpen
  \bibfield  {author} {\bibinfo {author} {\bibfnamefont {Alexander}\
  \bibnamefont {Gondarenko}}, \bibinfo {author} {\bibfnamefont {Stefan}\
  \bibnamefont {Preble}}, \bibinfo {author} {\bibfnamefont {Jacob}\
  \bibnamefont {Robinson}}, \bibinfo {author} {\bibfnamefont {Long}\
  \bibnamefont {Chen}}, \bibinfo {author} {\bibfnamefont {Hod}\ \bibnamefont
  {Lipson}}, \ and\ \bibinfo {author} {\bibfnamefont {Michal}\ \bibnamefont
  {Lipson}},\ }\bibfield  {title} {\enquote {\bibinfo {title} {Spontaneous
  {Emergence} of {Periodic} {Patterns} in a {Biologically} {Inspired}
  {Simulation} of {Photonic} {Structures}},}\ }\href {\doibase
  10.1103/PhysRevLett.96.143904} {\bibfield  {journal} {\bibinfo  {journal}
  {Phys. Rev. Lett.}\ }\textbf {\bibinfo {volume} {96}},\ \bibinfo {pages}
  {143904} (\bibinfo {year} {2006})}\BibitemShut {NoStop}%
\bibitem [{\citenamefont {Milano}\ and\ \citenamefont
  {Koumoutsakos}(2002)}]{milano_clustering_2002}%
  \BibitemOpen
  \bibfield  {author} {\bibinfo {author} {\bibfnamefont {Michele}\ \bibnamefont
  {Milano}}\ and\ \bibinfo {author} {\bibfnamefont {Petros}\ \bibnamefont
  {Koumoutsakos}},\ }\bibfield  {title} {\enquote {\bibinfo {title} {A
  {Clustering} {Genetic} {Algorithm} for {Cylinder} {Drag} {Optimization}},}\
  }\href {\doibase 10.1006/jcph.2001.6882} {\bibfield  {journal} {\bibinfo
  {journal} {J Comput. Phys.}\ }\textbf {\bibinfo {volume} {175}},\ \bibinfo
  {pages} {79} (\bibinfo {year} {2002})}\BibitemShut {NoStop}%
\bibitem [{\citenamefont {Berman}\ and\ \citenamefont
  {Wang}(2007)}]{berman_energy-minimizing_2007}%
  \BibitemOpen
  \bibfield  {author} {\bibinfo {author} {\bibfnamefont {Gordon~J.}\
  \bibnamefont {Berman}}\ and\ \bibinfo {author} {\bibfnamefont {Z.~Jane}\
  \bibnamefont {Wang}},\ }\bibfield  {title} {{\selectlanguage
  {english}\enquote {\bibinfo {title} {Energy-minimizing kinematics in hovering
  insect flight},}\ }}\href {\doibase 10.1017/S0022112007006209} {\bibfield
  {journal} {\bibinfo  {journal} {J. Fluid Mech.}\ }\textbf {\bibinfo {volume}
  {582}},\ \bibinfo {pages} {153} (\bibinfo {year} {2007})}\BibitemShut
  {NoStop}%
\bibitem [{\citenamefont {Weinacht}\ \emph {et~al.}(1998)\citenamefont
  {Weinacht}, \citenamefont {Ahn},\ and\ \citenamefont
  {Bucksbaum}}]{weinacht_measurement_1998}%
  \BibitemOpen
  \bibfield  {author} {\bibinfo {author} {\bibfnamefont {T.~C.}\ \bibnamefont
  {Weinacht}}, \bibinfo {author} {\bibfnamefont {J.}~\bibnamefont {Ahn}}, \
  and\ \bibinfo {author} {\bibfnamefont {P.~H.}\ \bibnamefont {Bucksbaum}},\
  }\bibfield  {title} {\enquote {\bibinfo {title} {Measurement of the
  {Amplitude} and {Phase} of a {Sculpted} {Rydberg} {Wave} {Packet}},}\ }\href
  {\doibase 10.1103/PhysRevLett.80.5508} {\bibfield  {journal} {\bibinfo
  {journal} {Phys. Rev. Lett.}\ }\textbf {\bibinfo {volume} {80}},\ \bibinfo
  {pages} {5508} (\bibinfo {year} {1998})}\BibitemShut {NoStop}%
\bibitem [{\citenamefont {Pisharody}\ and\ \citenamefont
  {Jones}(2002)}]{pisharody_phase-controlled_2002}%
  \BibitemOpen
  \bibfield  {author} {\bibinfo {author} {\bibfnamefont {S.~N.}\ \bibnamefont
  {Pisharody}}\ and\ \bibinfo {author} {\bibfnamefont {R.~R.}\ \bibnamefont
  {Jones}},\ }\bibfield  {title} {\enquote {\bibinfo {title} {Phase-controlled
  stair-step decay of autoionizing radial wave packets},}\ }\href {\doibase
  10.1103/PhysRevA.65.033418} {\bibfield  {journal} {\bibinfo  {journal} {Phys.
  Rev. A}\ }\textbf {\bibinfo {volume} {65}},\ \bibinfo {pages} {033418}
  (\bibinfo {year} {2002})}\BibitemShut {NoStop}%
\bibitem [{\citenamefont {Verlet}\ \emph {et~al.}(2002)\citenamefont {Verlet},
  \citenamefont {Stavros}, \citenamefont {Minns},\ and\ \citenamefont
  {Fielding}}]{verlet_controlling_2002}%
  \BibitemOpen
  \bibfield  {author} {\bibinfo {author} {\bibfnamefont {J.~R.~R.}\
  \bibnamefont {Verlet}}, \bibinfo {author} {\bibfnamefont {V.~G.}\
  \bibnamefont {Stavros}}, \bibinfo {author} {\bibfnamefont {R.~S.}\
  \bibnamefont {Minns}}, \ and\ \bibinfo {author} {\bibfnamefont {H.~H.}\
  \bibnamefont {Fielding}},\ }\bibfield  {title} {\enquote {\bibinfo {title}
  {Controlling the {Angular} {Momentum} {Composition} of a {Rydberg} {Electron}
  {Wave} {Packet}},}\ }\href {\doibase 10.1103/PhysRevLett.89.263004}
  {\bibfield  {journal} {\bibinfo  {journal} {Phys. Rev. Lett.}\ }\textbf
  {\bibinfo {volume} {89}},\ \bibinfo {pages} {263004} (\bibinfo {year}
  {2002})}\BibitemShut {NoStop}%
\bibitem [{\citenamefont {Noel}\ and\ \citenamefont
  {Stroud}(1997)}]{noel_shaping_1997}%
  \BibitemOpen
  \bibfield  {author} {\bibinfo {author} {\bibfnamefont {Michael~W.}\
  \bibnamefont {Noel}}\ and\ \bibinfo {author} {\bibfnamefont {C.~R.}\
  \bibnamefont {Stroud}},\ }\bibfield  {title} {\enquote {\bibinfo {title}
  {Shaping an atomic electron wave packet},}\ }\href {\doibase
  10.1364/OE.1.000176} {\bibfield  {journal} {\bibinfo  {journal} {Opt.
  Express}\ }\textbf {\bibinfo {volume} {1}},\ \bibinfo {pages} {176} (\bibinfo
  {year} {1997})}\BibitemShut {NoStop}%
\bibitem [{\citenamefont {Arb{\'o}}\ \emph {et~al.}(2003)\citenamefont
  {Arb{\'o}}, \citenamefont {Reinhold}, \citenamefont {Burgd{\"o}rfer},
  \citenamefont {Pattanayak}, \citenamefont {Stokely}, \citenamefont {Zhao},
  \citenamefont {Lancaster},\ and\ \citenamefont
  {Dunning}}]{arbo_pulse-induced_2003}%
  \BibitemOpen
  \bibfield  {author} {\bibinfo {author} {\bibfnamefont {D.~G.}\ \bibnamefont
  {Arb{\'o}}}, \bibinfo {author} {\bibfnamefont {C.~O.}\ \bibnamefont
  {Reinhold}}, \bibinfo {author} {\bibfnamefont {J.}~\bibnamefont
  {Burgd{\"o}rfer}}, \bibinfo {author} {\bibfnamefont {A.~K.}\ \bibnamefont
  {Pattanayak}}, \bibinfo {author} {\bibfnamefont {C.~L.}\ \bibnamefont
  {Stokely}}, \bibinfo {author} {\bibfnamefont {W.}~\bibnamefont {Zhao}},
  \bibinfo {author} {\bibfnamefont {J.~C.}\ \bibnamefont {Lancaster}}, \ and\
  \bibinfo {author} {\bibfnamefont {F.~B.}\ \bibnamefont {Dunning}},\
  }\bibfield  {title} {\enquote {\bibinfo {title} {Pulse-induced focusing of
  {Rydberg} wave packets},}\ }\href {\doibase 10.1103/PhysRevA.67.063401}
  {\bibfield  {journal} {\bibinfo  {journal} {Phys. Rev. A}\ }\textbf {\bibinfo
  {volume} {67}},\ \bibinfo {pages} {063401} (\bibinfo {year}
  {2003})}\BibitemShut {NoStop}%
\bibitem [{\citenamefont {Mestayer}\ \emph {et~al.}(2007)\citenamefont
  {Mestayer}, \citenamefont {Zhao}, \citenamefont {Lancaster}, \citenamefont
  {Dunning}, \citenamefont {Reinhold}, \citenamefont {Yoshida},\ and\
  \citenamefont {Burgd{\"o}rfer}}]{mestayer_transporting_2007}%
  \BibitemOpen
  \bibfield  {author} {\bibinfo {author} {\bibfnamefont {J.~J.}\ \bibnamefont
  {Mestayer}}, \bibinfo {author} {\bibfnamefont {W.}~\bibnamefont {Zhao}},
  \bibinfo {author} {\bibfnamefont {J.~C.}\ \bibnamefont {Lancaster}}, \bibinfo
  {author} {\bibfnamefont {F.~B.}\ \bibnamefont {Dunning}}, \bibinfo {author}
  {\bibfnamefont {C.~O.}\ \bibnamefont {Reinhold}}, \bibinfo {author}
  {\bibfnamefont {S.}~\bibnamefont {Yoshida}}, \ and\ \bibinfo {author}
  {\bibfnamefont {J.}~\bibnamefont {Burgd{\"o}rfer}},\ }\bibfield  {title}
  {\enquote {\bibinfo {title} {Transporting {Rydberg} {Electron} {Wave}
  {Packets} with {Chirped} {Trains} of {Pulses}},}\ }\href {\doibase
  10.1103/PhysRevLett.99.183003} {\bibfield  {journal} {\bibinfo  {journal}
  {Phys. Rev. Lett.}\ }\textbf {\bibinfo {volume} {99}},\ \bibinfo {pages}
  {183003} (\bibinfo {year} {2007})}\BibitemShut {NoStop}%
\bibitem [{\citenamefont {Maeda}\ \emph {et~al.}(2005)\citenamefont {Maeda},
  \citenamefont {Norum},\ and\ \citenamefont
  {Gallagher}}]{maeda_microwave_2005}%
  \BibitemOpen
  \bibfield  {author} {\bibinfo {author} {\bibfnamefont {H.}~\bibnamefont
  {Maeda}}, \bibinfo {author} {\bibfnamefont {D.~V.~L.}\ \bibnamefont {Norum}},
  \ and\ \bibinfo {author} {\bibfnamefont {T.~F.}\ \bibnamefont {Gallagher}},\
  }\bibfield  {title} {\enquote {\bibinfo {title} {Microwave {Manipulation} of
  an {Atomic} {Electron} in a {Classical} {Orbit}},}\ }\href {\doibase
  10.1126/science.1108470} {\bibfield  {journal} {\bibinfo  {journal}
  {Science}\ }\textbf {\bibinfo {volume} {307}},\ \bibinfo {pages} {1757}
  (\bibinfo {year} {2005})}\BibitemShut {NoStop}%
\bibitem [{\citenamefont {Feynman}\ \emph {et~al.}(2015)\citenamefont
  {Feynman}, \citenamefont {Hollingsworth}, \citenamefont {Vennettilli},
  \citenamefont {Budner}, \citenamefont {Zmiewski}, \citenamefont {Fahey},
  \citenamefont {Carroll},\ and\ \citenamefont {Noel}}]{feynman_quantum_2015}%
  \BibitemOpen
  \bibfield  {author} {\bibinfo {author} {\bibfnamefont {Rachel}\ \bibnamefont
  {Feynman}}, \bibinfo {author} {\bibfnamefont {Jacob}\ \bibnamefont
  {Hollingsworth}}, \bibinfo {author} {\bibfnamefont {Michael}\ \bibnamefont
  {Vennettilli}}, \bibinfo {author} {\bibfnamefont {Tamas}\ \bibnamefont
  {Budner}}, \bibinfo {author} {\bibfnamefont {Ryan}\ \bibnamefont {Zmiewski}},
  \bibinfo {author} {\bibfnamefont {Donald~P.}\ \bibnamefont {Fahey}}, \bibinfo
  {author} {\bibfnamefont {Thomas~J.}\ \bibnamefont {Carroll}}, \ and\ \bibinfo
  {author} {\bibfnamefont {Michael~W.}\ \bibnamefont {Noel}},\ }\bibfield
  {title} {\enquote {\bibinfo {title} {Quantum interference in the field
  ionization of {Rydberg} atoms},}\ }\href {\doibase
  10.1103/PhysRevA.92.043412} {\bibfield  {journal} {\bibinfo  {journal} {Phys.
  Rev. A}\ }\textbf {\bibinfo {volume} {92}},\ \bibinfo {pages} {043412}
  (\bibinfo {year} {2015})}\BibitemShut {NoStop}%
\bibitem [{\citenamefont {Gallagher}(1994)}]{gallagher_rydberg_1994}%
  \BibitemOpen
  \bibfield  {author} {\bibinfo {author} {\bibfnamefont {Thomas~F.}\
  \bibnamefont {Gallagher}},\ }\href@noop {} {{\selectlanguage {english}\emph
  {\bibinfo {title} {Rydberg {Atoms}}}}}\ (\bibinfo  {publisher} {Cambridge
  University Press},\ \bibinfo {address} {Cambridge ; New York},\ \bibinfo
  {year} {1994})\BibitemShut {NoStop}%
\bibitem [{\citenamefont {Jeys}\ \emph {et~al.}(1980)\citenamefont {Jeys},
  \citenamefont {Foltz}, \citenamefont {Smith}, \citenamefont {Beiting},
  \citenamefont {Kellert}, \citenamefont {Dunning},\ and\ \citenamefont
  {Stebbings}}]{jeys_diabatic_1980}%
  \BibitemOpen
  \bibfield  {author} {\bibinfo {author} {\bibfnamefont {T.~H.}\ \bibnamefont
  {Jeys}}, \bibinfo {author} {\bibfnamefont {G.~W.}\ \bibnamefont {Foltz}},
  \bibinfo {author} {\bibfnamefont {K.~A.}\ \bibnamefont {Smith}}, \bibinfo
  {author} {\bibfnamefont {E.~J.}\ \bibnamefont {Beiting}}, \bibinfo {author}
  {\bibfnamefont {F.~G.}\ \bibnamefont {Kellert}}, \bibinfo {author}
  {\bibfnamefont {F.~B.}\ \bibnamefont {Dunning}}, \ and\ \bibinfo {author}
  {\bibfnamefont {R.~F.}\ \bibnamefont {Stebbings}},\ }\bibfield  {title}
  {\enquote {\bibinfo {title} {Diabatic {Field} {Ionization} of {Highly}
  {Excited} {Sodium} {Atoms}},}\ }\href {\doibase 10.1103/PhysRevLett.44.390}
  {\bibfield  {journal} {\bibinfo  {journal} {Phys. Rev. Lett.}\ }\textbf
  {\bibinfo {volume} {44}},\ \bibinfo {pages} {390} (\bibinfo {year}
  {1980})}\BibitemShut {NoStop}%
\bibitem [{\citenamefont {F{\o}rre}\ and\ \citenamefont
  {Hansen}(2003)}]{forre_selective-field-ionization_2003}%
  \BibitemOpen
  \bibfield  {author} {\bibinfo {author} {\bibfnamefont {M.}~\bibnamefont
  {F{\o}rre}}\ and\ \bibinfo {author} {\bibfnamefont {J.~P.}\ \bibnamefont
  {Hansen}},\ }\bibfield  {title} {\enquote {\bibinfo {title}
  {Selective-field-ionization dynamics of a lithium $m=2$ {Rydberg} state:
  {Landau}-{Zener} model versus quantal approach},}\ }\href {\doibase
  10.1103/PhysRevA.67.053402} {\bibfield  {journal} {\bibinfo  {journal} {Phys.
  Rev. A}\ }\textbf {\bibinfo {volume} {67}},\ \bibinfo {pages} {053402}
  (\bibinfo {year} {2003})}\BibitemShut {NoStop}%
\bibitem [{\citenamefont {Zener}(1932)}]{zener_non-adiabatic_1932}%
  \BibitemOpen
  \bibfield  {author} {\bibinfo {author} {\bibfnamefont {C.}~\bibnamefont
  {Zener}},\ }\bibfield  {title} {{\selectlanguage {english}\enquote {\bibinfo
  {title} {Non-adiabatic crossing of energy levels},}\ }}\href {\doibase
  10.1098/rspa.1932.0165} {\bibfield  {journal} {\bibinfo  {journal} {Proc. R.
  soc. Lond. Ser. A-Contain. Pap. Math. Phys. Character}\ }\textbf {\bibinfo
  {volume} {137}},\ \bibinfo {pages} {696} (\bibinfo {year}
  {1932})}\BibitemShut {NoStop}%
\bibitem [{\citenamefont {Landau}(1932)}]{landau_zur_1932}%
  \BibitemOpen
  \bibfield  {author} {\bibinfo {author} {\bibfnamefont {Lev~D.}\ \bibnamefont
  {Landau}},\ }\bibfield  {title} {\enquote {\bibinfo {title} {Zur theorie der
  energieubertragung. {II}},}\ }\href
  {http://scholar.google.com/scholar?cluster=3232156340544517858&hl=en&oi=scholarr}
  {\bibfield  {journal} {\bibinfo  {journal} {Physics of the Soviet Union}\
  }\textbf {\bibinfo {volume} {2}},\ \bibinfo {pages} {28} (\bibinfo {year}
  {1932})}\BibitemShut {NoStop}%
\bibitem [{\citenamefont {Stueckelberg}(1932)}]{stueckelberg_theorie_1932}%
  \BibitemOpen
  \bibfield  {author} {\bibinfo {author} {\bibfnamefont {E.C.G.}\ \bibnamefont
  {Stueckelberg}},\ }\bibfield  {title} {\enquote {\bibinfo {title} {Theorie
  der unelastischen {St{\"o}sse} zwischen {Atomen}},}\ }\href {\doibase
  10.5169/seals-110177} {\bibfield  {journal} {\bibinfo  {journal} {Helv. Phys.
  Acta}\ }\textbf {\bibinfo {volume} {5}},\ \bibinfo {pages} {369} (\bibinfo
  {year} {1932})}\BibitemShut {NoStop}%
\bibitem [{\citenamefont {Majorana}(1932)}]{majorana_atomi_1932}%
  \BibitemOpen
  \bibfield  {author} {\bibinfo {author} {\bibfnamefont {Ettore}\ \bibnamefont
  {Majorana}},\ }\bibfield  {title} {\enquote {\bibinfo {title} {Atomi
  orientati in campo magnetico variabile},}\ }\href {\doibase
  10.1007/BF02960953} {\bibfield  {journal} {\bibinfo  {journal} {Il Nuovo
  Cimento}\ }\textbf {\bibinfo {volume} {9}},\ \bibinfo {pages} {43} (\bibinfo
  {year} {1932})}\BibitemShut {NoStop}%
\bibitem [{\citenamefont {Tada}\ \emph {et~al.}(2002)\citenamefont {Tada},
  \citenamefont {Kishimoto}, \citenamefont {Shibata}, \citenamefont {Kominato},
  \citenamefont {Yamada}, \citenamefont {Haseyama}, \citenamefont {Ogawa},
  \citenamefont {Funahashi}, \citenamefont {Yamamoto},\ and\ \citenamefont
  {Matsuki}}]{tada_manipulating_2002}%
  \BibitemOpen
  \bibfield  {author} {\bibinfo {author} {\bibfnamefont {M}~\bibnamefont
  {Tada}}, \bibinfo {author} {\bibfnamefont {Y}~\bibnamefont {Kishimoto}},
  \bibinfo {author} {\bibfnamefont {M}~\bibnamefont {Shibata}}, \bibinfo
  {author} {\bibfnamefont {K}~\bibnamefont {Kominato}}, \bibinfo {author}
  {\bibfnamefont {S}~\bibnamefont {Yamada}}, \bibinfo {author} {\bibfnamefont
  {T}~\bibnamefont {Haseyama}}, \bibinfo {author} {\bibfnamefont
  {I}~\bibnamefont {Ogawa}}, \bibinfo {author} {\bibfnamefont {H}~\bibnamefont
  {Funahashi}}, \bibinfo {author} {\bibfnamefont {K}~\bibnamefont {Yamamoto}},
  \ and\ \bibinfo {author} {\bibfnamefont {S}~\bibnamefont {Matsuki}},\
  }\bibfield  {title} {\enquote {\bibinfo {title} {Manipulating ionization path
  in a {Stark} map: {Stringent} schemes for the selective field ionization in
  highly excited {Rb} {Rydberg}},}\ }\href {\doibase
  10.1016/S0375-9601(02)01263-X} {\bibfield  {journal} {\bibinfo  {journal}
  {Phys. Lett. A}\ }\textbf {\bibinfo {volume} {303}},\ \bibinfo {pages} {285}
  (\bibinfo {year} {2002})}\BibitemShut {NoStop}%
\bibitem [{\citenamefont {G{\"u}rtler}\ and\ \citenamefont {van~der
  Zande}(2004)}]{gurtler_-state_2004}%
  \BibitemOpen
  \bibfield  {author} {\bibinfo {author} {\bibfnamefont {A.}~\bibnamefont
  {G{\"u}rtler}}\ and\ \bibinfo {author} {\bibfnamefont {W.~J.}\ \bibnamefont
  {van~der Zande}},\ }\bibfield  {title} {\enquote {\bibinfo {title}
  {$\ell$-state selective field ionization of rubidium {Rydberg} states},}\
  }\href {\doibase 10.1016/j.physleta.2004.02.062} {\bibfield  {journal}
  {\bibinfo  {journal} {Phys. Lett. A}\ }\textbf {\bibinfo {volume} {324}},\
  \bibinfo {pages} {315} (\bibinfo {year} {2004})}\BibitemShut {NoStop}%
\bibitem [{\citenamefont {Mitchell}(1998)}]{mitchell_introduction_1998}%
  \BibitemOpen
  \bibfield  {author} {\bibinfo {author} {\bibfnamefont {Melanie}\ \bibnamefont
  {Mitchell}},\ }\href@noop {} {{\selectlanguage {english}\emph {\bibinfo
  {title} {An {Introduction} to {Genetic} {Algorithms}}}}},\ \bibinfo {edition}
  {reprint edition}\ ed.\ (\bibinfo  {publisher} {A Bradford Book},\ \bibinfo
  {address} {Cambridge, Mass.},\ \bibinfo {year} {1998})\BibitemShut {NoStop}%
\bibitem [{\citenamefont {Goldberg}(1989)}]{goldberg_genetic_1989}%
  \BibitemOpen
  \bibfield  {author} {\bibinfo {author} {\bibfnamefont {David~E.}\
  \bibnamefont {Goldberg}},\ }\href@noop {} {{\selectlanguage {english}\emph
  {\bibinfo {title} {Genetic {Algorithms} in {Search}, {Optimization}, and
  {Machine} {Learning}}}}},\ \bibinfo {edition} {1st}\ ed.\ (\bibinfo
  {publisher} {Addison-Wesley Professional},\ \bibinfo {address} {Reading,
  Mass},\ \bibinfo {year} {1989})\BibitemShut {NoStop}%
\bibitem [{\citenamefont {Fahey}\ and\ \citenamefont
  {Noel}(2011)}]{fahey_excitation_2011}%
  \BibitemOpen
  \bibfield  {author} {\bibinfo {author} {\bibfnamefont {Donald~P.}\
  \bibnamefont {Fahey}}\ and\ \bibinfo {author} {\bibfnamefont {Michael~W.}\
  \bibnamefont {Noel}},\ }\bibfield  {title} {{\selectlanguage
  {english}\enquote {\bibinfo {title} {Excitation of {Rydberg} states in
  rubidium with near infrared diode lasers},}\ }}\href {\doibase
  10.1364/OE.19.017002} {\bibfield  {journal} {\bibinfo  {journal} {Opt.
  Express}\ }\textbf {\bibinfo {volume} {19}},\ \bibinfo {pages} {17002}
  (\bibinfo {year} {2011})}\BibitemShut {NoStop}%
\bibitem [{\citenamefont {Fahey}\ \emph {et~al.}(2015)\citenamefont {Fahey},
  \citenamefont {Carroll},\ and\ \citenamefont {Noel}}]{fahey_imaging_2015}%
  \BibitemOpen
  \bibfield  {author} {\bibinfo {author} {\bibfnamefont {Donald~P.}\
  \bibnamefont {Fahey}}, \bibinfo {author} {\bibfnamefont {Thomas~J.}\
  \bibnamefont {Carroll}}, \ and\ \bibinfo {author} {\bibfnamefont
  {Michael~W.}\ \bibnamefont {Noel}},\ }\bibfield  {title} {\enquote {\bibinfo
  {title} {Imaging the dipole-dipole energy exchange between ultracold rubidium
  {Rydberg} atoms},}\ }\href {\doibase 10.1103/PhysRevA.91.062702} {\bibfield
  {journal} {\bibinfo  {journal} {Phys. Rev. A}\ }\textbf {\bibinfo {volume}
  {91}},\ \bibinfo {pages} {062702} (\bibinfo {year} {2015})}\BibitemShut
  {NoStop}%
\end{thebibliography}%

\end{document}